\documentclass[aps,prl,twocolumn,superscriptaddress,showpacs]{revtex4}

\usepackage{graphicx}

\begin{document}



\title{Metallic ``Ferroelectricity'' in the Pyrochlore Cd$_2$Re$_2$O$_7$}

\author{I. A. Sergienko}
\affiliation{Dept. of Physics and Physical Oceanography, Memorial
University of Newfoundland, St. John's, NL, A1B 3X7, Canada}

\author{V. Keppens}
\altaffiliation{Permanent address: Dept. of Materials Science and
Engineering, The University of Tennessee, Knoxville, TN 37996}
\affiliation{Dept. of Physics and The National Center for Physical
Acoustics, The University of Mississippi, University, MS 38677}

\author{M. McGuire}
\altaffiliation{Present address: Dept. of Physics, Cornell
University, Ithaca, NY 14853} \affiliation{Dept. of Physics and
The National Center for Physical Acoustics, The University of
Mississippi, University, MS 38677}

\author{R. Jin}
\affiliation{Condensed Matter Sciences Division, Oak Ridge
National Lab., Oak Ridge, TN 37831}

\author{J. He}
\affiliation{Dept. of Physics, The University of Tennessee,
Knoxville, TN 37996}

\author{S. H. Curnoe}
\affiliation{Dept. of Physics and Physical Oceanography, Memorial
University of Newfoundland, St. John's, NL, A1B 3X7, Canada}

\author{B. C. Sales}
\affiliation{Condensed Matter Sciences Division, Oak Ridge
National Lab., Oak Ridge, TN 37831}

\author{P. Blaha}
\affiliation{Institute for Materials Chemistry, TU Vienna, A-1060
Vienna, Austria}

\author{D. J. Singh}
\affiliation{Code 6391, Naval Research Laboratory, Washington, DC
20375}

\author{K. Schwarz}
\affiliation{Institute for Materials Chemistry, TU Vienna, A-1060
Vienna, Austria}

\author{D. Mandrus}
\email[Corresponding author: ]{mandrusdg@ornl.gov}
\affiliation{Condensed Matter Sciences Division, Oak Ridge
National Lab., Oak Ridge, TN 37831}

\date{12 September 2003}

\begin{abstract}
A class of materials known as ``ferroelectric metals'' was
discussed theoretically by Anderson and Blount in 1965 [Phys. Rev.
Lett. \textbf{14}, 217 (1965)], but to date no examples of this
class have been reported. Here we present measurements of the
elastic moduli of Cd$_2$Re$_2$O$_7$ through the 200 K
cubic-to-tetragonal phase transition.   A Landau analysis of the
moduli reveals that the transition is consistent with
Cd$_2$Re$_2$O$_7$ being classified as a ``ferroelectric metal'' in
the weaker sense described by Anderson and Blount (loss of a
center of symmetry). First-principles calculations of the lattice
instabilities indicate that the dominant lattice instability
corresponds to a two-fold degenerate mode with $E_{u}$ symmetry,
and that motions of the O ions forming the O octahedra dominate
the energetics of the transition.

\end{abstract}


\maketitle

Itinerant electrons screen electric fields and inhibit the
electrostatic forces responsible for ferroelectric distortions.
Therefore, in a metallic system one does not expect to find
structural transitions similar to those found in insulating
materials with a tendency toward ferroelectricity.  The idea that
metallic behavior and ferroelectricity may not always be
incompatible had an early champion in B. T. Matthias
\cite{matthias}.  A groundbreaking paper on this subject was
written by P. W. Anderson and E. I. Blount (A\&B) in 1965
\cite{anderson}.  Applying Landau theory to a continuous
cubic-to-tetragonal (C-T) structural phase transition, A\&B
concluded that ``a transition from cubic to tetragonal in which
the only order parameter in Landau's sense is the unit cell shape,
{\it i.e.}, the strain, can be second order only with probability
zero.'' Applying these ideas to the C-T transition found in A-15
superconductors such as V$_3$Si and Nb$_3$Sn, A\&B concluded that
``these and perhaps several other metallic transitions may be
`ferroelectric' in the sense of the appearance of a polar axis, or
possibly at least involve the loss of a center of symmetry.''
Ultimately, however, it was shown that the structural transitions
in the A-15 compounds were not continuous but rather weakly first
order, and that strain was indeed the appropriate order parameter
\cite{testardi}. As no other materials seemed to fit A\&B's
criteria, ideas regarding metallic ``ferroelectricity'' have not
been pursued over the past thirty-eight years.

The pyrochlore Cd$_2$Re$_2$O$_7$ has attracted attention recently
as an oxide superconductor on a geometrically frustrated lattice
\cite{hanawa, sakai, jin, vyaselev, lumsden}.  The normal state
properties of Cd$_2$Re$_2$O$_7$ are also intriguing \cite{jin2002,
huo, hiroi, arai, eguchi, wang, sakai2002}, particularly the C-T
transition at 200 K that profoundly affects the electronic
structure, transport, and magnetic susceptibility of this
material. Evidence from resistivity, specific heat, NQR, and X-ray
diffraction \cite{jin2002, arai, sakai2002, castellan, yamaura}
indicates that the 200 K C-T transition is continuous. Another,
first-order, structural phase transition ($T_C$ = 120 K ) has been
reported in Cd$_2$Re$_2$O$_7$ \cite{hiroi2002}, but because the
issues raised by this transition are secondary to the main focus
of this paper, we will only mention this lower-temperature
transition briefly in what follows.  Despite several studies
\cite{sakai2002, castellan, yamaura}, the structure of
Cd$_2$Re$_2$O$_7$ below 200 K has not been fully determined,
mainly because the departure from cubic symmetry is extremely
small and even high-resolution X-ray measurements \cite{castellan,
yamaura} can barely detect the splitting of the cubic Bragg peaks.
It has been shown, however, that there is no multiplication of the
unit cell below the 200 K transition (the transition is
ferrodistortive), and that a loss of three-fold symmetry
accompanies the transition \cite{vyaselev}. There is also some
evidence that inversion symmetry may be broken in the
low-temperature phases \cite{yamaura}. Cd$_2$Re$_2$O$_7$,
therefore, is a good candidate for becoming the first material to
obey A\&B's criteria provided that strain can be ruled out as the
primary order parameter for the C-T transition at 200 K.

Single crystals of Cd$_2$Re$_2$O$_7$ were grown from the vapor in
sealed silica tubes using Cd metal (5N, Johnson Matthey) and
Re$_2$O$_7$ (3N, Johnson Matthey) \cite{jin, jin2002}.  The
measurements reported in this paper were performed on a crystal
cut into a rectangular parallelepiped with dimensions 1.2 x 1.9 x
2.3 mm$^3$. The sample was oriented with all faces perpendicular
to the crystallographic $<100>$ axes. The experimental density of
the sample was 8.795 g/cm$^3$; this can be compared with the x-ray
density of 8.814 g/cm$^3$.

Resonant Ultrasound Spectroscopy (RUS) measurements were performed
as a function of temperature (5-300 K) to determine the elastic
moduli of the sample. RUS is a technique developed by Migliori,
{\it et al.} \cite{migliori} for determining the complete elastic
tensor of a small single crystal by measuring its free-body
resonances. This method has the advantage that all moduli can be
determined simultaneously, thereby avoiding remounts of
transducers and multiple temperature sweeps.

\begin{figure}
\includegraphics[width = 3 in] {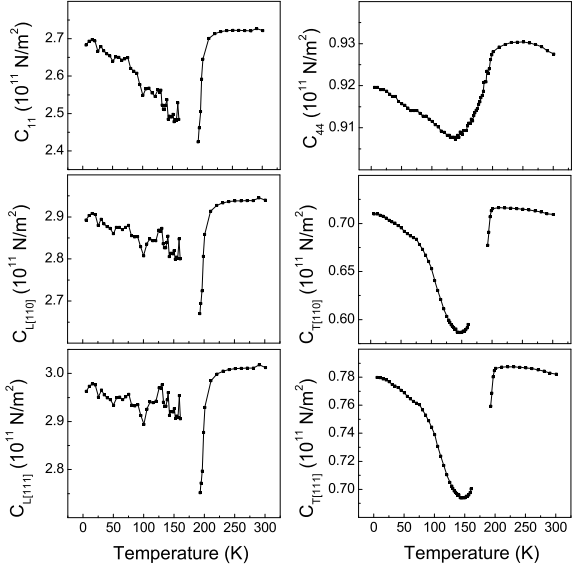}
\caption{Elastic moduli vs. temperature obtained on a single
crystal of Cd$_2$Re$_2$O$_7$ using RUS. The relationship between
sound velocity and elastic moduli is given by $v_s =
(C/\rho)^{1/2}$, where $C$ is the effective elastic constant and
$\rho$ is the density. In a cubic crystal, longitudinal waves
propagating in the [100] direction are governed by $C_{11}$, and
both transverse waves are governed by $C_{44}$. In the [110]
direction, longitudinal waves are governed by $C_{L[110]}$ =
${1/2}(C_{11} + C_{12} + 2C_{44})$, one transverse wave is
governed by $C_{44}$, and the other by $C_{T[110]} = {1/2}(C_{11}
- C_{12})$. In the [111] direction, longitudinal waves are
governed by $C_{L[111]} = {1/3}(C_{11} + 2C_{12} + 4C_{44})$, and
both transverse waves are governed by $C_{T[111]} = {1/3}(C_{11} -
C_{12} + C_{44})$.  Note the small magnitude (2\%) of the anomaly
in $C_{44}$ compared to the anomalies in the other moduli.}
\end{figure}

In Fig. 1 we plot the elastic moduli of Cd$_2$Re$_2$O$_7$ vs.
temperature for the three modes of elastic waves in the principal
propagation directions in a cubic system \cite{kittel}.  Deep into
the transition the ultrasonic absorption of the sample became so
great that for several temperatures not enough resonances were
observed to allow for an accurate determination of all three
elastic moduli.  However, the lowest frequency resonance depended
almost exclusively on $C_{44}$ and was visible throughout the
entire transition.

If strain were the order parameter, we would expect at least one
of these elastic constants to soften dramatically as expected for
an elastic instability.  This is not observed.  The salient
feature of Fig. 1 is the step-like change at 200 K in all of the
moduli except $C_{44}$.  This behavior of the moduli can be
modeled using the order parameter proposed by Sergienko and Curnoe
\cite{sergienko}. This order parameter involves collective atomic
displacements corresponding to a long-wavelength phonon of $E_u$
symmetry.

The minimal model for Landau free energy $F$ which accounts for
the anomalies of the elastic moduli should include a
ferrodistortive energy $F_d$ expanded in terms of the structural
order parameter $(\eta_1,\eta_2)$, the elastic energy $F_{el}$,
and coupling between the order parameter and strain $F_{d-el}$:
\begin{widetext}
\begin{equation}
\begin{array}{lcll}
F_d&=&a_1(\eta_1^2+\eta_2^2)+a_2(\eta_1^2+\eta_2^2)^2+a_3(\eta_1^2+\eta_2^2)^3
+b_1(\eta_1^3-3\eta_1\eta_2^2)^2, \vspace{3pt}\\ F_{el}&=&1/2
C_{11}^0(e_1^2+e_2^2+e_3^2)+C_{12}^0(e_1e_2+e_2e_3+e_1e_3) +1/2
C_{44}^0(e_4^2+e_5^2+e_6^2), \vspace{3pt}\\ F_{d-el}&=&\lambda_1
(\eta_1^2+\eta_2^2)(e_1+e_2+e_3)+\lambda_2[(\eta_1^2-\eta_2^2)(e_1+e_2-2e_3)
+2\sqrt3 \eta_1\eta_2(e_1-e_2)]\vspace{3pt}\\ &&
\quad+\mu_1(\eta_1^2+\eta_2^2)(e_4^2+e_5^2+e_6^2)+\mu_2[(\eta_1^2-\eta_2^2)(e_4^2+e_5^2-2e_6^2)
+2\sqrt3 \eta_1\eta_2(e_4^2-e_5^2)],
\end{array}
\end{equation}
\end{widetext}
with $F=F_d+F_{el}+F_{d-el}$.  Here $C_{11}^0, C_{12}^0$, and
$C_{44}^0$ are the elastic moduli in the cubic phase
$(\eta_1=\eta_2=0)$ and $a_1, a_2, a_3$, and $b_1$ are the Landau
expansion coefficients, $e_i$ is the strain, and $\lambda_i$ and
$\mu_i$ are coupling constants.

Terms up to fourth order in $F_d$ are isotropic, therefore sixth
order terms are required to lift the degeneracy between ordered
states. Since the phase transition at $T_C=200$~K is second order,
the coefficient $a_2$ must be positive. The elastic moduli in the
tetragonal phase ($\eta_1=0$, $\eta_2\ne0$) can be calculated
using Slonczewski-Thomas formalism \cite{slonczewski} $$
C_{ij}=\frac{\partial^2 F}{\partial e_i\partial e_j} -
\frac{\partial^2 F}{\partial \eta_2\partial e_i} \frac{\partial^2
F}{\partial \eta_2\partial e_j} \left(\frac{\partial^2 F}{\partial
\eta_2^2}\right)^{-1}, $$ where the equibrium values of $\eta_2$
and the strain tensor are calculated from the system of equations

\begin{equation}
 \frac{\partial F}{\partial \eta_2} = \frac{\partial
F}{\partial e_i} = 0, \quad i=1,\ldots, 6.
\end{equation}
We obtain
\begin{equation}
\begin{array}{l}
C_{11} = C_{11}^0 - A^2/2a_2 + O(\eta_2^2)\\ C_{33} = C_{11}^0 -
B^2/2a_2 + O(\eta_2^2)\\ C_{12} = C_{12}^0 - A^2/2a_2 +
O(\eta_2^2)\\ C_{13} = C_{12}^0 - AB/2a_2 + O(\eta_2^2)\\ C_{44} =
C_{44}^0 + O(\eta_2^2)\\ C_{66} = C_{66}^0 + O(\eta_2^2),
\end{array}
\end{equation}
where $A = \lambda_1 - \lambda_2$, $B = \lambda_1 + 2\lambda_2$.
Higher order terms can easily be calculated from the above
equations by expanding in $\eta_2$ but the resulting expressions
are cumbersome. Steps are therefore expected in the elastic moduli
$C_{11}$, $C_{33}$, $C_{12}$, and $C_{13}$ at the continuous
transition, while the shear moduli $C_{44}$ and $C_{66}$ have
continuous anomalies.  The data in Fig. 1 are qualitatively
consistent with these predictions \cite{note1}.

In many respects the elastic behavior of Cd$_2$Re$_2$O$_7$ (CRO)
resembles that of SrTiO$_3$ (STO), but there are some important
differences. In STO the coupling is linear in all components of
the strain but quadratic in the order parameter \cite{migliori,
luthi, slonczewski, fossheim}, whereas in CRO it is linear only in
the diagonal components $e_1$, $e_2$, $e_3$ of the strain tensor.
It should also be kept in mind that STO is antiferrodistortive and
inversion symmetry is not broken in the tetragonal phase. Like in
STO, the measured elastic anomalies of $C_{11}$ and
$(C_{11}-C_{12})/2$ in CRO are not true step functions but are
broadened by several degrees. L\"{u}thi and Moran \cite{luthi}
ascribe this behavior in STO to residual strain, and there is no
reason not to expect this strain in CRO as well. The downward
``dip'' observed near $T_C$ in the longitudinal elastic moduli has
also been observed in STO, and can be ascribed to order parameter
fluctuations \cite{fossheim}. Domain formation below $T_C$
introduces several difficulties into a quantitative analysis of
the data, both because of the anisotropies associated with domains
and because domain wall motions make important contributions to
the elastic moduli. To obtain elastic moduli below the transition
we assumed a random domain distribution and that the sample
retained a macroscopic cubic symmetry. Resolving these
difficulties requires a detailed understanding of the
microstructure of Cd$_2$Re$_2$O$_7$, and this knowledge is not yet
available. However, none of these difficulties affects the
phenomenology presented here.

\begin{figure}
\includegraphics[width = 2 in, angle =-90]{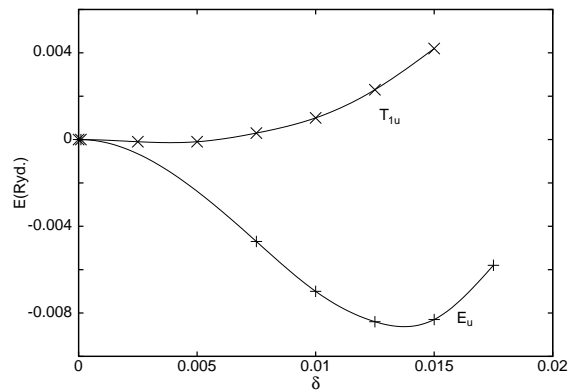}
\caption{Relativistic LDA energetics of lattice instabilities in
Cd$_2$Re$_2$O$_7$. The coordinates of the O(1) atoms are in
Cartesian coordinates, units of the lattice constant a = 10.219
\AA: (0.315- $\delta_E$ + $\delta_T$, 0.625, 0.625), (0.935 +
$\delta_E$ + $\delta_T$, 0.625, 0.625), (0.625, 0.315, 0.625),
(0.625, 0.935, 0.625), (0.625, 0.625, 0.315 + $\delta_E$), (0.625,
0.625, 0.935 - $\delta_E$), (0.375, 0.065, 0.375), (0.375, 0.685,
0.375), (0.065 + $\delta_E$ + $\delta_T$, 0.375, 0.375), (0.685 -
$\delta_E$ + $\delta_T$, 0.375, 0.375), (0.375, 0.375, 0.685 +
$\delta_E$), (0.375, 0.375, 0.065 - $\delta_E$), where $E$ and $T$
are the $E_u$ and $T_{1u}$ frozen-in amplitudes, respectively.}
\end{figure}

To help elucidate the origin of the lattice instabilities, first
principles calculations were performed in the local density
approximation (LDA) using the general potential linearized
augmented planewave (LAPW) method \cite{singh} as described in
Ref. \cite{singh94, blaha}. Initially, scalar relativistic
calculations of the atomic forces were performed for a sufficient
number of small atomic displacements away from the equilibrium
structure to determine the dynamical matrix for an 88 atom
supercell.  Then the full phonon dispersion relations were
obtained by a direct method using the PHONON program
\cite{parlinski}.  At the zone center two very unstable modes were
found, a two-fold degenerate $E_u$ symmetry mode, and a three-fold
degenerate $T_{1u}$ mode. Both of these modes involved breaking of
inversion symmetry and have eigenvectors that are heavily
dominated by the O(1) site (forming the O octahedra coordinating
the Re ions).  In addition, two more weakly unstable and several
low frequency but stable modes involving Cd and Re atom shifts
were found, many of them also breaking inversion symmetry.

Since the unstable modes were heavily dominated by O(1) motion, we
neglected the minor components and calculated the energetics as a
function of distortion amplitude with a tetragonal cell for the
$E_u$  and $T_{1u}$ displacement patterns.  These calculations
were done relativistically, including spin-orbit, which we find to
significantly affect the energetics by reducing the tendency
towards lattice instability. Nonetheless, we still find a
substantial instability of the $E_u$ mode and a marginal
instability of the $T_{1u}$ mode, as shown in Fig. 2.  Clearly,
the dominant instability corresponds to the $E_u$ mode, and so we
expect that the higher temperature phase transition is due to its
freezing in.

The lower temperature transition might be associated with a change
of symmetry into which this mode freezes (note that it is two-fold
degenerate), as in, e.g., the
cubic-tetragonal-orthorhombic-rhombohedral transition sequence of
BaTiO$_3$ under cooling.  Freezing of the doubly degenerate $E_u$
mode alone may result in three possible low symmetry structures
$I\bar{4}m2$, $I4_122$, and $F222$ \cite{sergienko}. We have
performed the LDA calculations for these three space groups and
found that the energetics are nearly the same.  This can be
understood because terms up to fourth-order in $F_d$ are
isotropic. In Fig. 2 we show the calculations, corresponding to
the $I4_122$ structure, which was proposed to be the lowest
temperature phase in Ref. \cite{yamaura}. In this case, soft modes
associated with the $T_{1u}$ could play an important role in the
low-temperature superconductivity.

Alternately, we note that the two unstable modes are of different
symmetry, and therefore do not interact at lowest order.  Perhaps
a mode related to the $T_{1u}$ O(1) displacement but with
additional metal and/or O(2) displacement is unstable enough to
freeze in and give the lower temperature transition.

The temperature dependence of the elastic moduli shown in Fig. 1
allows us to rule out strain as an order parameter in
Cd$_2$Re$_2$O$_7$. The conclusions of Anderson and Blount
\cite{anderson} can therefore be applied to Cd$_2$Re$_2$O$_7$: the
order parameter must either be (1) ``some electronic mystery
parameter,'' or (2) ``some change in symmetry, such as the loss of
the inversion center.''  At present, the evidence supports the
second possibility with most likely candidate for the order
parameter being a small, coherent (hence ``ferroelectric'')
collective displacement of the atoms with $E_u$ symmetry,
dominated by the motion of the O(1) atom as discussed above.  Even
though, following A\&B, we use the term ``ferroelectric,'' we
would like to stress that although inversion symmetry is broken
there is no evidence that a polar axis is formed; indeed, our
analysis is fully consistent with the $I\bar{4}m2$ space group
proposed in Ref. \cite{yamaura}, and this space group is
piezoelectric, although because Cd$_2$Re$_2$O$_7$ is metallic, no
piezoelectric coupling is possible here.

It is an open question whether the concept of a ``ferroelectric
metal'' will be fruitful in explaining the anomalous physical
properties at the 200 K transition in Cd$_2$Re$_2$O$_7$. In many
ways, the 200 K transition is reminiscent of a charge density wave
(CDW) transition such as that observed in TaSe$_2$, although CDWs
in cubic materials such as pyrochlores are generally not expected.
Moreover, electronic structure calculations indicate a nearly
isotropic Fermi surface and no obvious nesting or CDW instability
\cite{singh}.  If Cd$_2$Re$_2$O$_7$ is indeed a ``ferroelectric
metal'' however, one can imagine a redistribution of charge within
the material and physical properties that mimic a CDW transition.
Also, the dramatic decrease of the electrical resistivity of
Cd$_2$Re$_2$O$_7$ below 200 K finds a natural explanation in terms
of reduced scattering from the unstable ions as they freeze in. It
is hoped that the identification of Cd$_2$Re$_2$O$_7$ as a
``ferroelectric metal'' will stimulate theoretical development of
the unique continuous phase transition found in this material.

\begin{acknowledgments}
I. S. and S. C. were supported by NSERC of Canada.  Work at UT was
supported by NSF DMR-007 2998 and at U. Miss. by NSF DMR-020 6625.
Oak Ridge National Laboratory is managed by UT-Battelle, LLC, for
the US Department of Energy under contract DE-AC05-00OR22725.
Computations were performed using facilities of the DoD HPCMO ASC
center.  Work at the Naval Research Laboratory is supported by the
Office of Naval Research.
\end{acknowledgments}

\bibliography{mf_final}

\begin{thebibliography}{29}
\expandafter\ifx\csname natexlab\endcsname\relax\def\natexlab#1{#1}\fi
\expandafter\ifx\csname bibnamefont\endcsname\relax
  \def\bibnamefont#1{#1}\fi
\expandafter\ifx\csname bibfnamefont\endcsname\relax
  \def\bibfnamefont#1{#1}\fi
\expandafter\ifx\csname citenamefont\endcsname\relax
  \def\citenamefont#1{#1}\fi
\expandafter\ifx\csname url\endcsname\relax
  \def\url#1{\texttt{#1}}\fi
\expandafter\ifx\csname urlprefix\endcsname\relax\def\urlprefix{URL }\fi
\providecommand{\bibinfo}[2]{#2}
\providecommand{\eprint}[2][]{\url{#2}}

\bibitem[{\citenamefont{Fisk}()}]{matthias}
\bibinfo{author}{\bibfnamefont{Z.}~\bibnamefont{Fisk}}, \bibinfo{note}{private
  communication.}

\bibitem[{\citenamefont{Anderson and Blount}(1965)}]{anderson}
\bibinfo{author}{\bibfnamefont{P.~W.} \bibnamefont{Anderson}} \bibnamefont{and}
  \bibinfo{author}{\bibfnamefont{E.~I.} \bibnamefont{Blount}},
  \bibinfo{journal}{Phys. Rev. Lett.} \textbf{\bibinfo{volume}{14}},
  \bibinfo{pages}{217} (\bibinfo{year}{1965}).

\bibitem[{\citenamefont{Testardi and Bateman}(1967)}]{testardi}
\bibinfo{author}{\bibfnamefont{L.~R.} \bibnamefont{Testardi}} \bibnamefont{and}
  \bibinfo{author}{\bibfnamefont{T.~B.} \bibnamefont{Bateman}},
  \bibinfo{journal}{Phys. Rev.} \textbf{\bibinfo{volume}{154}},
  \bibinfo{pages}{402} (\bibinfo{year}{1967}).

\bibitem[{\citenamefont{Hanawa et~al.}(2001)}]{hanawa}
\bibinfo{author}{\bibfnamefont{M.}~\bibnamefont{Hanawa}} \bibnamefont{et~al.},
  \bibinfo{journal}{Phys. Rev. Lett.} \textbf{\bibinfo{volume}{87}},
  \bibinfo{pages}{187001} (\bibinfo{year}{2001}).

\bibitem[{\citenamefont{Sakai et~al.}(2001)}]{sakai}
\bibinfo{author}{\bibfnamefont{H.}~\bibnamefont{Sakai}} \bibnamefont{et~al.},
  \bibinfo{journal}{J. Phys. Condens. Matter} \textbf{\bibinfo{volume}{13}},
  \bibinfo{pages}{L785} (\bibinfo{year}{2001}).

\bibitem[{\citenamefont{Jin et~al.}(2001)}]{jin}
\bibinfo{author}{\bibfnamefont{R.}~\bibnamefont{Jin}} \bibnamefont{et~al.},
  \bibinfo{journal}{Phys. Rev. B} \textbf{\bibinfo{volume}{64}},
  \bibinfo{pages}{180503} (\bibinfo{year}{2001}).

\bibitem[{\citenamefont{Vyaselev et~al.}(2002)}]{vyaselev}
\bibinfo{author}{\bibfnamefont{O.}~\bibnamefont{Vyaselev}}
  \bibnamefont{et~al.}, \bibinfo{journal}{Phys. Rev. Lett.}
  \textbf{\bibinfo{volume}{89}}, \bibinfo{pages}{17001} (\bibinfo{year}{2002}).

\bibitem[{\citenamefont{Lumsden et~al.}(2002)}]{lumsden}
\bibinfo{author}{\bibfnamefont{M.~D.} \bibnamefont{Lumsden}}
  \bibnamefont{et~al.}, \bibinfo{journal}{Phys. Rev. Lett.}
  \textbf{\bibinfo{volume}{89}}, \bibinfo{pages}{147002}
  (\bibinfo{year}{2002}).

\bibitem[{\citenamefont{Jin et~al.}(2002)}]{jin2002}
\bibinfo{author}{\bibfnamefont{R.}~\bibnamefont{Jin}} \bibnamefont{et~al.},
  \bibinfo{journal}{J. Phys. Condens. Matter} \textbf{\bibinfo{volume}{14}},
  \bibinfo{pages}{L117} (\bibinfo{year}{2002}).

\bibitem[{\citenamefont{Huo et~al.}(2002)}]{huo}
\bibinfo{author}{\bibfnamefont{D.}~\bibnamefont{Huo}} \bibnamefont{et~al.},
  \bibinfo{journal}{J. Phys. Condens. Matter} \textbf{\bibinfo{volume}{14}},
  \bibinfo{pages}{L257} (\bibinfo{year}{2002}).

\bibitem[{\citenamefont{Hiroi et~al.}(2002{\natexlab{a}})}]{hiroi}
\bibinfo{author}{\bibfnamefont{Z.}~\bibnamefont{Hiroi}} \bibnamefont{et~al.},
  \bibinfo{journal}{J. Phys. Soc. Jpn.} \textbf{\bibinfo{volume}{71}},
  \bibinfo{pages}{1553} (\bibinfo{year}{2002}{\natexlab{a}}).

\bibitem[{\citenamefont{Arai et~al.}(2002)}]{arai}
\bibinfo{author}{\bibfnamefont{K.}~\bibnamefont{Arai}} \bibnamefont{et~al.},
  \bibinfo{journal}{J. Phys. Condens. Matter} \textbf{\bibinfo{volume}{14}},
  \bibinfo{pages}{L461} (\bibinfo{year}{2002}).

\bibitem[{\citenamefont{Eguchi et~al.}(2002)}]{eguchi}
\bibinfo{author}{\bibfnamefont{R.}~\bibnamefont{Eguchi}} \bibnamefont{et~al.},
  \bibinfo{journal}{Phys. Rev. B} \textbf{\bibinfo{volume}{66}},
  \bibinfo{pages}{12516} (\bibinfo{year}{2002}).

\bibitem[{\citenamefont{Wang et~al.}(2002)}]{wang}
\bibinfo{author}{\bibfnamefont{N.~L.} \bibnamefont{Wang}} \bibnamefont{et~al.},
  \bibinfo{journal}{Phys. Rev. B} \textbf{\bibinfo{volume}{66}},
  \bibinfo{pages}{14534} (\bibinfo{year}{2002}).

\bibitem[{\citenamefont{Sakai et~al.}(2002)}]{sakai2002}
\bibinfo{author}{\bibfnamefont{H.}~\bibnamefont{Sakai}} \bibnamefont{et~al.},
  \bibinfo{journal}{Phys. Rev. B} \textbf{\bibinfo{volume}{66}},
  \bibinfo{pages}{100509} (\bibinfo{year}{2002}).

\bibitem[{\citenamefont{Castellan et~al.}(2002)}]{castellan}
\bibinfo{author}{\bibfnamefont{J.~P.} \bibnamefont{Castellan}}
  \bibnamefont{et~al.}, \bibinfo{journal}{Phys. Rev. B}
  \textbf{\bibinfo{volume}{66}}, \bibinfo{pages}{134528}
  (\bibinfo{year}{2002}).

\bibitem[{\citenamefont{Yamaura and Hiroi}(2002)}]{yamaura}
\bibinfo{author}{\bibfnamefont{J.-I.} \bibnamefont{Yamaura}} \bibnamefont{and}
  \bibinfo{author}{\bibfnamefont{Z.}~\bibnamefont{Hiroi}}, \bibinfo{journal}{J.
  Phys. Soc. Jpn.} \textbf{\bibinfo{volume}{71}}, \bibinfo{pages}{2598}
  (\bibinfo{year}{2002}).

\bibitem[{\citenamefont{Hiroi et~al.}(2002{\natexlab{b}})\citenamefont{Hiroi,
  Yamaura, Muraoka, and Hanawa}}]{hiroi2002}
\bibinfo{author}{\bibfnamefont{Z.}~\bibnamefont{Hiroi}},
  \bibinfo{author}{\bibfnamefont{J.-I.} \bibnamefont{Yamaura}},
  \bibinfo{author}{\bibfnamefont{Y.}~\bibnamefont{Muraoka}}, \bibnamefont{and}
  \bibinfo{author}{\bibfnamefont{M.}~\bibnamefont{Hanawa}},
  \bibinfo{journal}{J. Phys. Soc. Jpn.} \textbf{\bibinfo{volume}{71}},
  \bibinfo{pages}{1634} (\bibinfo{year}{2002}{\natexlab{b}}).

\bibitem[{\citenamefont{Migliori et~al.}(1993)}]{migliori}
\bibinfo{author}{\bibfnamefont{A.}~\bibnamefont{Migliori}}
  \bibnamefont{et~al.}, \bibinfo{journal}{Physica B}
  \textbf{\bibinfo{volume}{183}}, \bibinfo{pages}{1} (\bibinfo{year}{1993}).

\bibitem[{\citenamefont{Kittel}(1996)}]{kittel}
\bibinfo{author}{\bibfnamefont{C.}~\bibnamefont{Kittel}},
  \emph{\bibinfo{title}{Introduction to Solid State Physics, Seventh Edition}}
  (\bibinfo{publisher}{John Wiley and Sons}, \bibinfo{year}{1996}).

\bibitem[{\citenamefont{Sergienko and Curnoe}(2003)}]{sergienko}
\bibinfo{author}{\bibfnamefont{I.~A.} \bibnamefont{Sergienko}}
  \bibnamefont{and} \bibinfo{author}{\bibfnamefont{S.~H.}
  \bibnamefont{Curnoe}}, \bibinfo{journal}{J. Phys. Soc. Jpn.}
  \textbf{\bibinfo{volume}{72}}, \bibinfo{pages}{1607} (\bibinfo{year}{2003}).

\bibitem[{\citenamefont{Slonczewski and Thomas}(1970)}]{slonczewski}
\bibinfo{author}{\bibfnamefont{J.~C.} \bibnamefont{Slonczewski}}
  \bibnamefont{and} \bibinfo{author}{\bibfnamefont{H.}~\bibnamefont{Thomas}},
  \bibinfo{journal}{Phys. Rev. B} \textbf{\bibinfo{volume}{1}},
  \bibinfo{pages}{3599} (\bibinfo{year}{1970}).

\bibitem[{not()}]{note1}
\bibinfo{note}{Assuming all domains occur with equal probability, one can
  calculate effective cubic elastic constants below T$_C$ by averaging over
  different domains. Thus a longitudinal wave propagating along the $<100>$
  direction has a 1/3 contribution from $C_{33}$ and a 2/3 contribution from
  $C_{11}$, etc. See Ref. 24 for details of this calculation applied to
  SrTiO$_{3}$.}

\bibitem[{\citenamefont{L{\"u}thi and Moran}(1970)}]{luthi}
\bibinfo{author}{\bibfnamefont{B.}~\bibnamefont{L{\"u}thi}} \bibnamefont{and}
  \bibinfo{author}{\bibfnamefont{T.~J.} \bibnamefont{Moran}},
  \bibinfo{journal}{Phys. Rev. B} \textbf{\bibinfo{volume}{2}},
  \bibinfo{pages}{1211} (\bibinfo{year}{1970}).

\bibitem[{\citenamefont{Fossheim and Berre}(1972)}]{fossheim}
\bibinfo{author}{\bibfnamefont{K.}~\bibnamefont{Fossheim}} \bibnamefont{and}
  \bibinfo{author}{\bibfnamefont{B.}~\bibnamefont{Berre}},
  \bibinfo{journal}{Phys. Rev. B} \textbf{\bibinfo{volume}{5}},
  \bibinfo{pages}{3292} (\bibinfo{year}{1972}).

\bibitem[{\citenamefont{Singh et~al.}(2002)\citenamefont{Singh, Blaha, Schwarz,
  and Sofo}}]{singh}
\bibinfo{author}{\bibfnamefont{D.~J.} \bibnamefont{Singh}},
  \bibinfo{author}{\bibfnamefont{P.}~\bibnamefont{Blaha}},
  \bibinfo{author}{\bibfnamefont{K.}~\bibnamefont{Schwarz}}, \bibnamefont{and}
  \bibinfo{author}{\bibfnamefont{J.~O.} \bibnamefont{Sofo}},
  \bibinfo{journal}{Phys. Rev. B} \textbf{\bibinfo{volume}{65}},
  \bibinfo{pages}{155109} (\bibinfo{year}{2002}).

\bibitem[{\citenamefont{Singh}(1994)}]{singh94}
\bibinfo{author}{\bibfnamefont{D.~J.} \bibnamefont{Singh}},
  \emph{\bibinfo{title}{Planewaves, Pseudopotentials and the LAPW Method}}
  (\bibinfo{publisher}{Kluwer Academic, Boston}, \bibinfo{year}{1994}).

\bibitem[{\citenamefont{Blaha et~al.}(2001)\citenamefont{Blaha, Schwarz,
  Madsen, Kvasnicka, and Luitz}}]{blaha}
\bibinfo{author}{\bibfnamefont{P.}~\bibnamefont{Blaha}},
  \bibinfo{author}{\bibfnamefont{K.}~\bibnamefont{Schwarz}},
  \bibinfo{author}{\bibfnamefont{G.}~\bibnamefont{Madsen}},
  \bibinfo{author}{\bibfnamefont{D.}~\bibnamefont{Kvasnicka}},
  \bibnamefont{and} \bibinfo{author}{\bibfnamefont{J.}~\bibnamefont{Luitz}},
  \emph{\bibinfo{title}{WIEN2k, An augmented plane wave + local orbitals
  program for calculating crystal properties}} (\bibinfo{publisher}{K.Schwarz,
  TU Vienna, Austria 2001, ISBN 3-9501031-1-2}, \bibinfo{year}{2001}).

\bibitem[{\citenamefont{Parlinski}()}]{parlinski}
\bibinfo{author}{\bibfnamefont{K.}~\bibnamefont{Parlinski}},
  \bibinfo{note}{software PHONON (2002)}.

\end{thebibliography}

\end{document}